\documentclass[twocolumn,showkeys,showpacs,preprintnumbers,prd,superscriptaddress,nofootinbib]{revtex4}
\usepackage{graphicx}
\usepackage{epsf}
\usepackage{bm}
\usepackage{amsmath}
\usepackage{amsfonts}
\usepackage{amssymb}
\usepackage{epstopdf}
\usepackage{natbib}
\usepackage{hyperref}
\usepackage{color}
\usepackage{verbatim}
\usepackage{multirow}
\usepackage{bm}
\usepackage{hyperref}
\usepackage{listings}
\usepackage{soul}

\usepackage[T1]{fontenc}
\usepackage{array}

\begin{document}

\title{A Joint Analysis of Strong Lensing and Type Ia Supernovae to
Determine the Hubble Constant}

\author{L. R. Cola\c{c}o}
\email{colacolrc@gmail.com}
\affiliation{Universidade Federal do Rio Grande do Norte, Departamento de F\'{i}sica Te\'{o}rica e Experimental, 59300-000, Natal - RN, Brazil.}

\author{R. F. L. Holanda}
\email{holandarfl@gmail.com}
\affiliation{Universidade Federal do Rio Grande do Norte, Departamento de F\'{i}sica Te\'{o}rica e Experimental, 59300-000, Natal - RN, Brazil.}
\affiliation{Departamento de F\'{\i}sica, Universidade Federal de Campina Grande, 58429-900, Campina Grande - PB, Brasil}

\author{Z. C. Santana}
\email{zilmarjunior@hotmail.com.br}
\affiliation{Universidade Federal do Rio Grande do Norte, Departamento de F\'{i}sica Te\'{o}rica e Experimental, 59300-000, Natal - RN, Brazil.}

\author{R. Silva}
\email{raimundosilva@fisica.ufrn.br}
\affiliation{Universidade Federal do Rio Grande do Norte, Departamento de F\'{i}sica Te\'{o}rica e Experimental, 59300-000, Natal - RN, Brazil.}

\begin{abstract}

We present a cosmological model-independent determination of the Hubble constant, $H_0$, by combining time-delay measurements from seven TDCOSMO systems, Einstein radius measurements, and Type Ia Supernovae data sourced from the Pantheon+ sample. For each lens of time-delay system, we calculate the angular diameter distance $D_{A_l}$ using the product $D^{\textrm{Obs}}(z_l) \cdot D_{A,\Delta t}^{\textrm{Obs}}(z_l, z_s)$, where $D^{\textrm{Obs}}(z_l)$ is reconstructed via Gaussian Processes from 99 Einstein radius measurements, and $D_{A,\Delta t}^{\textrm{Obs}}(z_l,z_s)$ is the time-delay angular distance. We also reconstruct the unanchored luminosity distance $H_0 D_L(z_l)$ from supernova data. By using the cosmic distance duality relation validity, we anchor $D_{A_l}$ and $H_0 D_L(z_l)$ to infer $H_0 = 70.55 \pm 7.44$ km/s/Mpc (68\% CL). Our result, though not resolving the Hubble tension, offers a cosmological model-independent consistency check and highlights the potential of using strong lensing and supernovae data via the cosmic distance duality relation to constrain $H_0$.

\end{abstract}

\maketitle

\section{Introduction}

The discovery of accelerated expansion of the universe \cite{Riess1998,Perlmutter1999,Weinberg2013} has presented significant theoretical challenges, particularly in understanding its mechanisms. The standard model of cosmology attributes this acceleration to dark energy. In contrast, dark matter governs large-scale structure formation \cite{zer1986,Sahni2004,CalderaCabral2009}, yielding predictions consistent with observations such as the Planck CMB data \cite{2016}. However, notable tensions remain, including discrepancies in the Hubble constant ($H_0$),  measurements between early- and late-universe methods \cite{Kamionkowski2023,DiValentino:2021izs}, the cosmological constant problem \cite{Weinberg1989,Padmanabhan2003,Lombriser2023}, and the cosmic coincidence problem \cite{Zlatev1999}. Locally, $H_0=73.04 \pm 1.04$ km/s/Mpc \cite{Riess2021jrx}, while Planck estimates $H_0=67.36 \pm 0.54$  km/s/Mpc \cite{Planck2018vyg}, a $\sim 5\sigma$ tension identified as a key issue in modern cosmology \cite{Freedman2017}. Additional discrepancies, such as those in $\sigma_8$ \cite{Battye2015}, have prompted alternative models involving modified gravity \cite{Pogosian2022}, particle physics \cite{Alcaniz2022,DEramo2018}, thermodynamics \cite{daSilva2021}, among others \cite{DiValentino:2021izs}. Lensed quasars also offer information on $H_0$ tensions \cite{Wong2019}.

The powerful technique known as strong lensing time-delay cosmography (TDC) offers a direct measurement of the Hubble constant that is independent of the local distance ladder or probes anchored to sound-horizon physics \cite{1964MNRAS128307R,Birrer2022chj,Saha2024axf,2024SSRv22087S,Moresco2022phi}. Advances in photometric precision have led to the discovery of numerous multiply-imaged quasars and supernovae \cite{2018MNRAS4811041T,2023MNRAS5203305L,Kelly2014mwa,Rodney2021keu}, as well as precise time-delay measurements \cite{Millon2020ugy,2017arXiv170609424C}. In particular, the H0LiCOW and SHARP Collaborations analyzed six individual lenses and achieved $H_0$ measurements with precision ranging from approximately $4.3\%$ to $9.1\%$ \cite{suyu711,Suyu2013kha,101093mnrasstw3077,2019MNRAS4844726B,101093mnrasstz3451,mnrasstz2547}. The STRIDES Collaboration measured $H_0$ with a precision of approximately $3.9\%$ also using TDC \cite{mnrasstaa828}. Despite these developments, all measurements follow a standard procedure and incorporate single-aperture stellar kinematics for each lens \cite{101093mnrasstx483}. The H0LiCOW Collaboration later reported $H_0 = 73.2 \pm 1.75$ km/s/Mpc with $2.4\%$ precision \cite{Wong2019}, further refined to $2\%$ by including blind measurements \cite{Millon2019slk}. By parameterizing lens-type galaxies with a power-law or stars+dark-matter mass profile, high accuracy was achieved, while relaxing these assumptions yielded $8\%$ precision, consistent with both \textit{Planck} and SH0ES results \cite{Birrer2020tax}.

A suitable application of time-delay galaxy lenses for inferring $H_0$ involves combining with reliable distance indicators from other probes using the distance sum rule \cite{Collett2019hrr}. Such an approach has been successfully implemented with type Ia supernovae (SNe Ia) \cite{Collett2019hrr}, gamma-ray bursts (GRBs) \cite{Du2023zsz}, multi-messenger gravitational wave standard sirens \cite{Liao2019hfl}, ultra-compact radio sources \cite{2021MNRAS5032179Q}, quasars (QSOs) \cite{2020ApJ897127W}, and others. Recently, \cite{Gong2024yne} proposed an inverse distance ladder technique to determine $H_0$ by only assuming the validity of the cosmic distance-duality relation \cite{Renzi2020fnx}. Using angular diameter distances of time-delay lenses and supernovae observations, the authors reported $H_0$ in good agreement with SH0ES measurements within $\approx 1.3\sigma$. Furthermore, \cite{Li2023gpp} determined $H_0$ in a model-independent manner using Gaussian process (GP) regression on mock quasars and time-delay distances, finding $H_0 = 70.8 \pm 1.5$ km/s/Mpc with $5\%$ uncertainty in distance values (see \cite{Colaco2023gzy,Gonzalez2024qjs,Liao2020zko,Perivolaropoulos2024yxv} for more works that use the CDDR validity with other observational data to determine $H_0$).

A recent work uses the inverse distance ladder technique to determine $H_0$ by combining unanchored luminosity distances from the Pantheon$+$ sample with angular diameter distances from galaxy clusters obtained from their Sunyaev-Zel’dovich effect (SZE) and X-ray observations \cite{Colaco2023gzy}. Under minimal cosmological assumptions, the authors reported $H_0 = 67.22 \pm 6.07$ km / s / Mpc at the confidence level $1\sigma$, with an error margin of approximately $9\%$. On the other hand, by using two galaxy cluster gas mass fraction measurement samples, unanchored distances from SNe Ia, and the validity of the cosmic distance duality relation, the authors in \cite{Gonzalez2024qjs} reported another estimate of $H_0$ that is also independent of any specific cosmological framework. Their joint analysis yielded $H_0 = 73.4 \pm 5.95$ km/s/Mpc at $68\%$ confidence level. In \cite{Colaco2025aqp}, relative distances from the SNe Ia observations were anchored to absolute distance measurements from the time-delay of SGL systems to estimate $H_0$. Such analysis yielded $H_0 = 75.57 \pm 4.415$ km/s/Mpc at $1\sigma$, indicating consistency with late-universe probes.

This paper aims to constrain the Hubble constant using a combination of independent datasets by assuming the validity of the cosmic distance-duality relation. We propose anchoring relative luminosity distances from SNe Ia with angular diameter distances obtained from a combination of Time-delay and Einstein radius measurements of strong lensing systems. This approach provides a cosmological model-independent way of inferring $H_0$. To accomplish this, we used the most extensive compiled data set on SNe Ia to date, known as Pantheon$+$ \cite{Scolnic2021amr}, a data set of 7 two-image time-delay lensing systems from TDCOSMO Collaboration \cite{Birrer2020tax}, and 99 selected SGL systems from \cite{Cao2015qja}. 

The paper is organized as follows: Section \ref{Methodology} introduces the methodology employed and the datasets used to perform the statistical analyses. We then discuss the statistical construction of our observables. Our findings regarding the constraints on $H_0$ are addressed in Section \ref{rasults}, followed by the final remarks in Section \ref{final}.

\section{Data and Methodology}
\label{Methodology}

The main approach to constrain $H_0$ relies on the validity of the CDDR: 

\begin{equation}\label{CDDR}
D_L = (1+z)^2 D_A (z),
\end{equation}
where $D_L$ and $D_A$ represent the luminosity and angular diameter distances, respectively. This relation holds under the assumption that the photon number remains conserved along null geodesics in a Riemannian space-time between the source and the observer. Due to current technological advances, several observational datasets have allowed for the accurate investigation of a possible deviation of the CDDR. However, no statistically significant evidence has been found up to this point \cite{Wang2024rxm,Favale2024sdq,Kumar2021djt,Lima2021slf,Holanda2021vvh,Goncalves2019xtc,Holanda2019vmh,Yang2017bkv,Rana2017sfr}, and its broad applicability underscores the fundamental importance in observational cosmology. Any deviation from it could signal the presence of new physics or systematic errors in observations \cite{Ellis2007,CDDR,Bassett2003vu}. Following \cite{Renzi2020fnx}, it is possible to write

\begin{equation}\label{H0NA}
    H_0 = \frac{1}{(1+z)^2}\frac{ \Theta^{\textrm{SNe}}(z) }{D_A (z)},
\end{equation}
where $\Theta^{\textrm{SNe}}(z) \equiv [H_0 D_L]^{\textrm{SNe}} (z)$ represents the unanchored luminosity distance. This relation highlights the possibility of deriving estimates of $H_0$ if we have the unanchored luminosity distance and the angular diameter distance measurements at the same redshift $z$. It is important to note that this relationship is based on the assumption that the validity of the CDDR means that all distance probes consistently trace the cosmic expansion. If this assumption were not valid, the value of $H_0$ would not remain constant; rather, it would exhibit an unphysical trend with redshift. This highlights the role of $H_0$ not only as an absolute distance scale but also as a factor that reveals inconsistencies among distance probes. 

We utilize Type Ia Supernovae (SNe Ia) to derive $\Theta (z)$. Additionally, we combine time-delay distances with Einstein radius measurements of SGL systems to derive $D_A (z)$, both at the lens redshift ($z_l$), and hence to estimate $H_0$. More details will be provided as follows.

\subsection{Einstein Radius - $\theta_E$ }

An Einstein ring is a purely gravitational phenomenon that occurs when the source ($s$), the lens ($l$), and the observer ($o$) are in the same line of sight forming a ring-like structure with an angular radius $\theta_E$ (Einstein radius) \cite{Cao2015qja}. In general, quasars act as a background source at redshift $z_s$, which is lensed by foreground elliptical galaxies at $z_l$, and multiple bright images of the active galactic nuclei are formed around their host galaxies. Furthermore, the characteristic distances used to describe the physics of an SGL involve only angular diameter distances (ADD), and the lens model generally fitted to the observed images is based on the singular isothermal ellipsoid model \cite{1999AJ1172010R}\ footnote {The projected mass distribution is elliptical.}. However, the approach employed here takes a more straightforward methodology and assumes spherical symmetry. Thus, under the rigid assumption of the so-called Singular Isothermal Sphere (SIS) model, the Einstein radius is given by \cite{Cao2015qja}:

\begin{equation} \label{SIS}
    \theta_E = 4\pi\frac{\sigma_{SIS}^{2}}{c^2}\frac{D_{A_{ls}}}{D_{A_s}},
\end{equation}
where $D_{A_{ls}}$ is the ADD between the lens and the source, and $D_{A_s}$ is the ADD between the observer and the source. The quantities $c$ and $\sigma_{SIS}$ are the speed of light and the velocity dispersion measured under the SIS model assumption, respectively.    

However, recent studies using SGL systems have demonstrated that the slopes of density profiles for individual galaxies significantly deviate from the SIS model. This indicates that the SIS model does not accurately represent the lens mass distribution \cite{2009ApJ703L51K,2010ApJ724511A,2011MNRAS4152215B,Sonnenfeld2013xga,Cao2015qja,Cao2016wor,Chen2018jcf}. Therefore, in line with more recent works, we adopt the power-law model (PLAW) for the mass distribution of lensing systems. This model assumes a spherically symmetric mass distribution characterized by a more general power-law index ($\gamma$), defined as $\rho \propto r^{-\gamma}$, where $\rho$ represents the total mass distribution and $r$ is the spherical radius of the lensing galaxy center. By assuming that velocity anisotropy can be neglected and applying the spherical Jeans equation \cite{Ofek2003sp}, it is possible to rescale the dynamical mass inside the aperture of size $\theta_{ap}$ projected to the lens plane and obtain the following observational quantity \cite{Cao2015qja}:

\begin{equation}\label{DSIS}
    D^{\rm{Obs}} \equiv \frac{D_{A_{ls}}}{D_{A_s}} = \frac{c^2 \theta_E}{4\pi \sigma_{\rm{ap}}^{2}} \Bigg( \frac{\theta_{\rm{ap}}}{\theta_E} \Bigg)^{2-\gamma}f^{-1}(\gamma), 
\end{equation}
where $\sigma_{ap}$ is the stellar velocity dispersion inside the aperture $\theta_{ap}$, and    

\begin{eqnarray}
    f(\gamma) & \equiv & -\frac{1}{\sqrt{\pi}} \frac{(5-2\gamma)(1-\gamma)}{3-\gamma}\frac{\Gamma (\gamma-1)}{\Gamma (\gamma -3/2)} \nonumber \\
     && \times \left[     \frac{\Gamma (\gamma/2 - 1/2)}{\Gamma (\gamma/2)}\right]^2.
\end{eqnarray}
If $\gamma=2$, we recover the SIS model. In this paper, we choose $\gamma=2.1$ \cite{Ofek2003sp,Colaco2020wbn}. Note that the observational quantity \ref{DSIS} is independent of the Hubble constant value; it gets canceled in the distance ratio $D_{A_{ls}}/D_{A_s}$.

As argued in the Ref. \cite{Cao2015qja}, for a single system, one may use the line-of-sight velocity dispersion ($\sigma_{\rm{ap}}$), but since we are dealing with a sample, it is necessary to convert all velocity dispersions measured within an aperture into velocity dispersions within circular aperture of radius $R{\rm{eff}}/2$ following the description \cite{1995MNRAS2761341J}: $\sigma_0= \sigma_{\rm{ap}}( \theta_{\rm{eff}}/ 2\theta_{\rm{ap}}))^{-0.04}$, where $\theta_{\rm{eff}}$ represents the effective angular radius. While using $\sigma_{\rm{ap}}$ is consistent with the model, adopting $\sigma_0$ makes the observable $D^{\rm{Obs}}$ more homogeneous for a set of lenses located at different redshifts. Therefore, following \cite{Cao2015qja}, we replace $\sigma_{\rm{ap}}$ with $\sigma_0$ in Eq. \ref{DSIS} (see Table I in such reference).

We concentrate on a specific catalog that contains 118 confirmed sources of SGL systems and are presented in Table 1 of \cite{Cao2015qja}. This catalog includes systems sourced from various surveys, including the SLOAN Lens ACS, the BOSS Emission-line Lens Survey (BELLS), the LSD, and the Strong Legacy Survey SL2S. All redshifts have been determined spectroscopically, and the Einstein radii were measured by fitting pixelized images of the sources. To minimize the overall scatter in the data, it is essential to exclude the systems with $D^{\textrm{Obs}}>1$ \cite{Colaco2020ndf,Colaco2020wbn,Colaco2022aut,Cao2018rzc,Amante2019xao,Chen2018jcf,Cao2016wor,Holanda2021mcn,Colaco2022noc}. Non-physical values of $D^{\textrm{Obs}}$ arise when applying the PLAW model. Since the distance $D_{A_s}$ is greater than $D_{A_{ls}}$, the ratio $D_{A_{ls}}/D_{A_s}$ in Eq. \ref{DSIS} must be less than unity for any SGL system. If measurements yield $D^{\textrm{Obs}} > 1$, it would indicate $D_{A_{ls}} > D_{A_s}$, which is not rational in the context of the SGL phenomenon. Furthermore, estimating any cosmological parameter considering systems with $D^{\rm{Obs}} > 1$ certainly leads to larger $\chi_{\textrm{red}}^2$. As a result, we use $99$ SGL systems in the redshift ranges of $  0.075 \leq z_l \leq 1.004$ and $0.196 \leq z_s \leq 3.595$.

Although selecting the SIS/PLAW model is fundamental, it is important to highlight that this approach may not be the most effective way to ensure an accurate mass profile. It is demonstrated in \cite{Martel2002qi} that the presence of background matter tends to increase the image separations produced by lensing galaxies, a finding supported by ray-tracing simulations in Cold Dark Matter (CDM) models\footnote{This effect is relatively small.}. Another study indicated that the richer environments of early-type galaxies may host a higher ratio of dwarf to giant galaxies compared to those found in the field \cite{2000ApJ545145C}. However, in \cite{Keeton2000pz}, this effect has been shown to nearly counterbalance the influence of background matter, rendering the distribution of image separations largely independent of the environment. On the other hand, it is predicted that the lenses in groups have a mean image separation of approximately $0.2$ arcseconds smaller than those found in the field. According to \cite{2012JCAP03016C}, all these factors can significantly influence image separation, which could affect the estimation of $D^{\textrm{Obs}}$ by as much as $\wp=20\%$.

Finally, following the methodology described by \cite{Grillo2007iv,Cao2015qja}, we assign uncertainties in Einstein radii as $\sigma_{\theta_E} = 0.05 \theta_E$ ($5\%$ for all systems). Thus, the uncertainty of Eq. \ref{DSIS} shall be \cite{Cao2015qja}:

\begin{equation}
\sigma_{D^{\textrm{Obs}}} = D^{\textrm{Obs}} \left\{4\Bigg(  \frac{\sigma_{\sigma_{0}}}{\sigma_{0}}  \Bigg)^2 + (1-\gamma)^2\Bigg( \frac{\sigma_{\theta_E}}{\theta_E}  \Bigg)^2 + \wp^2 \right\}^{1/2}.
\end{equation}
It is worth noting that some lensing systems produce two images while others produce four. This distinction could introduce systematic differences between the two sub-groups. However, previous analysis of \cite{2015AJ....149....2M} demonstrated no significant differences between two-image and four-image systems.

\subsection{Time-Delay Angular Diameter Distance - $D_{A,\Delta t}$}

The time-delay between multiple images of strongly lensed quasars has been extensively used to infer the Hubble constant \cite{2020MNRAS4981420W,Birrer2020tax,2020MNRAS4931725K,Pandey2019yic}. Based on the fact that photons propagate null geodesics and they come from a distant source with distinct optical paths, they must pass through distinct gravitational potentials \cite{1964MNRAS128307R,PhysRevLett13789,2010ARAA4887T}, and the lensing time-delay between any two images is determined by the geometry of the universe together with the gravity field of the lensing-type galaxy. For a two-image lens system ($A$ and $B$) with the SIS mass profile describing the lensing mass, it is possible to obtain:

\begin{equation}\label{Def_DADT}
    \Delta t=\Delta \tau (A) - \Delta \tau (B) = \frac{(1+z_l)}{2c}\frac{D_{A_l}D_{A_s}}{D_{A_{ls}}}[\theta_{A}^{2} - \theta_{B}^{2}].
\end{equation}
Defining the quantity $\frac{D_{A_l}D_{A_s}}{D_{A_{ls}}} \equiv D_{A,\Delta t}^{\textrm{Obs}}(z_l,z_s)$ as the time-delay angular diameter distance, or simply the time-delay distance, it is obtained:

\begin{equation}\label{DADt}
   D_{A, \Delta t}^{\textrm{Obs}}(z_l,z_s) \equiv  \frac{D_{A_l}D_{A_s}}{D_{A_{ls}}} = \frac{2c \Delta t}{(1+z_l)(\theta_{A}^{2} - \theta_{B}^{2})}.
\end{equation}

\begin{figure}[htbp]
\includegraphics[width=0.49\textwidth]{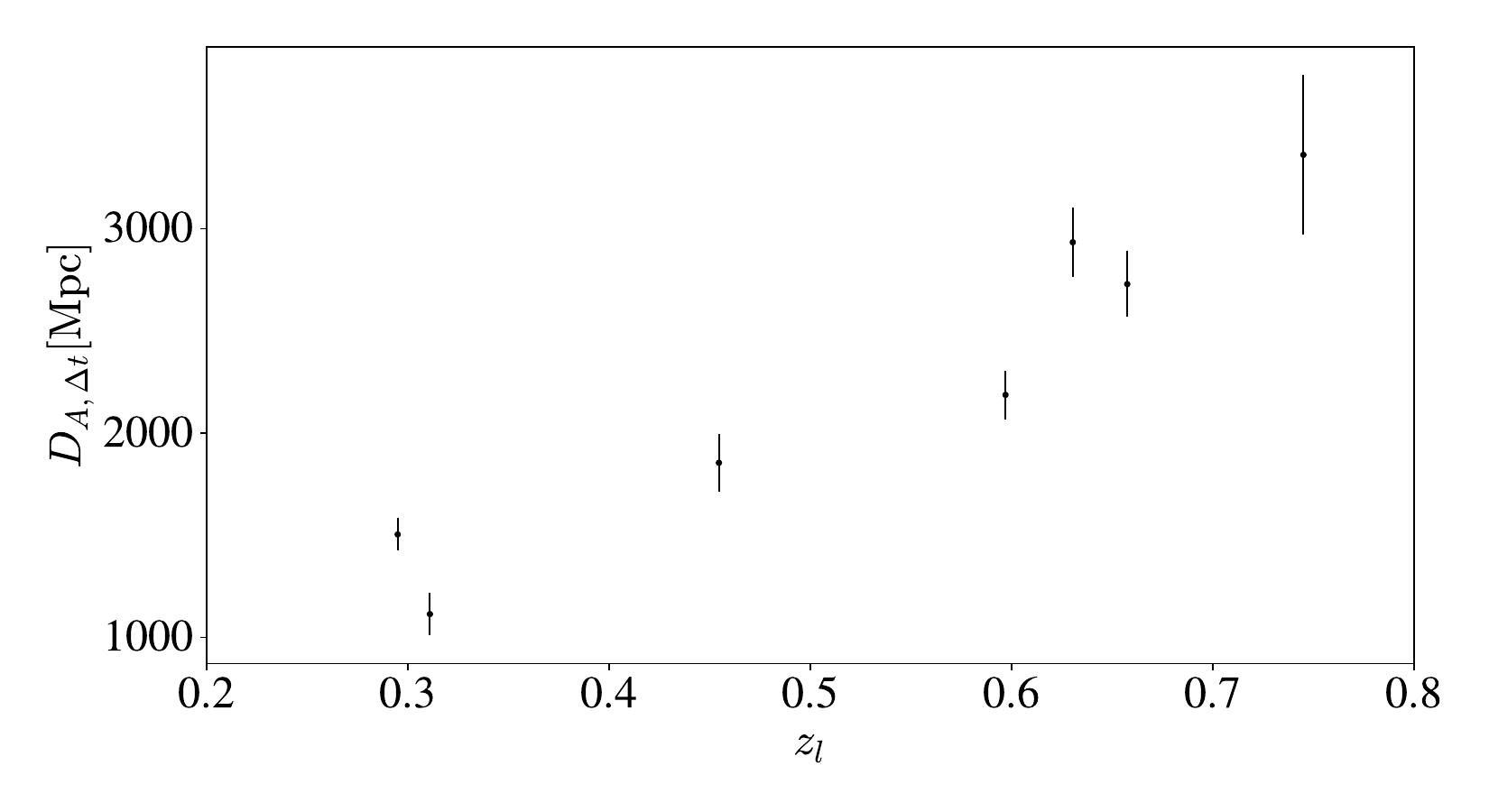}
\caption{The seven well-studied and refined time-delay distance estimates ($D_{A,\Delta t}^{\rm{Obs}} (z_l,z_s)$) regarding $z_l$, along with corresponding error bars from the TDCOSMO Collaboration \cite{Birrer2020tax}.}
\label{DADT_fig}
\end{figure}

For the two-image time-delay lensing systems described by Eq. \ref{DADt}, we consider seven well-studied SGL systems that have precise time-delay measurements between the lensed images. The TDCOSMO collaboration has released this data set in \cite{Birrer2020tax}, six of which are from the H0LiCOW collaboration \cite{2020MNRAS4981420W}\footnote{Available at http://www.h0licow.org.}. The posterior products for these systems were calculated using a hierarchical Bayesian approach, where the Mass-Sheet Transform (MST) is constrained solely by stellar kinematics. The redshifts of both the lens ($z_l$) and the source ($z_s$), along with their inferred time-delay distance ($D_{A,\Delta t}^{\rm{Obs}} (z_l,z_s)$) based on the Power-Law (PLAW) model, the half-light radius ($r_{eff}$), the Einstein radius ($\theta_E$), the power-law slope ($\gamma_{pl}$), and the external convergence ($\kappa_{\rm{ext}}$), are displayed in Table 2 of \cite{Birrer2020tax} (see Figure \ref{DADT_fig}). It is important to note that the observational quantity $D_{A, \Delta t}^{\textrm{Obs}}(z_l,z_s)$ depends solely on the lens model and is independent of the cosmological model \cite{Kelly2023mgv}.

This paper aims to obtain a cosmology model-independent estimate of $H_0$ by combining observables derived from strong lensing time-delay and Einstein radius measurements, along with Type Ia supernovae data. To achieve this, we combine the observed quantities $D_{A, \Delta t}^{\textrm{Obs}}(z_l, z_s)$  and $D^{\textrm{Obs}}(z_l)$  to calculate the angular diameter distance to the lens for each of the seven time-delay system, as follows:

\begin{equation}\label{DAl}
    D_{A_l} = D^{\textrm{Obs}}(z_l) \cdot D_{A,\Delta t}^{\textrm{Obs}}(z_l, z_s).
\end{equation}
The quantity $D^{\textrm{Obs}}(z_l)$ is reconstructed using Gaussian Process (GP) regression techniques (see Fig. \ref{GPs}, left panel) with data from Ref. \cite{Cao2015qja}. This process estimates its value at $z_l$ of each of the seven lenses in the time-delay systems. This procedure allows us to calculate the angular diameter distance to each lens based on Eq. \ref{DAl}. Subsequently, these values are anchored to the corresponding $\Theta^{\textrm{SNe}}(z_l)$, which is also reconstructed at the same lens redshifts using GP regression (see Fig. \ref{GPs}, right panel).

Finally, substituting these quantities into the following relation:

\begin{equation}\label{H0Obs}
    H_0^{\textrm{Est}} = \frac{1}{(1+z_l)^2} \cdot \frac{\Theta^{\textrm{SNe}}(z_l)}{D_{A_l}},
\end{equation}
we obtain seven cosmological model-independent estimates of the Hubble constant.

As one may see, we focus on two distinct samples of strong lensing systems: one where the Einstein radii were measured, and another where the time-delay distances were determined. Although the lens and source redshifts differ between these two samples, we confidently multiplied the lensing and time-delay measurements to estimate the angular diameter distance to the lenses in the time-delay systems. This approach is entirely justified, as for each lens redshift in the time-delay sample, there are multiple lensing systems with comparable lens redshifts and source redshifts clustered around those in time-delay systems (see \cite{Rana2017sfr}).

\subsection{The unanchored luminosity distances - $\Theta^{\textrm{SNe}} (z)$}

Observations of SNe Ia may also provide the so-called unanchored luminosity distance $\Theta^{\textrm{SNe}} (z)$ (see details in \cite{Renzi2020fnx}). The relative distances are derived from the apparent magnitude of SNe Ia using the relation:

\begin{equation}\label{SNemb}
    m_b = 5 \log_{10}[\Theta^{\textrm{SNe}}(z)] - 5a_B.
\end{equation}
The parameter $a_B$ represents the intercept of the SNe Ia magnitude-redshift relation.  As discussed in the Ref. \cite{Riess2016jrr}, this parameter exhibits a weak dependence on the cosmological model, as it can be obtained with a cosmographic expansion (e.g., in terms of $q_0$, $j_0$, etc.), thereby avoiding the need to assume a specific model such as $\Lambda$CDM. This approach is particularly useful in studies aiming to test the consistency of different cosmological models. The authors from the Ref. \cite{Riess2016jrr} found $a_B = 0.71273 \pm 0.00176$. Recently, the Ref. \cite{Riess2021jrx} reviewed these assumptions and found no signs of inconsistency with the findings of \cite{Riess2016jrr}.

We use one of the largest compiled datasets of SNe Ia, the so-called Pantheon$+$ \cite{Scolnic2021amr},  in order to obtain $\Theta ^{\textrm{SNe}}(z)$ at the same $z_l$ of the 7 time-delay systems considered in this study. For that purpose, we assume that the deflection of light occurs in the local Minkowski space-time of the lens, which is perturbed by its gravitational potential \cite{Narayan1996ba}. This implies that all physical quantities involved in the deviation of the light path correspond to their values at $z_l$. The Pantheon$+$ sample consists of 1701 light curves of 1550 different SNe Ia in a redshift range of $0.001 \leq z \leq 2.261$; it has a significant increase compared to the original Pantheon sample, especially at lower $z$. For our purposes, it is essential to convert the apparent magnitudes of this sample into a set of unanchored luminosity distances. Considering Eq. \ref{SNemb}) we may obtain:

\begin{equation}
\label{H0DL}
    \Theta^{\textrm{SNe}} (z) = 10^{(m_b(z) + 5a_B)/5} = 10^{m'_B(z)/5}, 
\end{equation}
where $m'_b \equiv m_b + 5a_B$. 

To estimate $\Theta^{\textrm{SNe}}(z)$ uncertainties accurately, including their correlations, we need to consider the covariance matrix of the apparent magnitudes, which encompasses both statistical and systematic uncertainties, along with the error in $a_B$\footnote{If we know that the covariance matrix is not diagonal but choose to set the off-diagonal elements to zero, this decision can lead inaccurate uncertainty estimates. Ignoring these correlations may result in an underestimation of the precision of the analysis or affect the best-fit parameters' central values.}. Considering the dependence of the covariance matrix of SNe Ia on cosmological parameters that are not known, we can estimate the errors on $H_0$ accurately. It improves the robustness of the resulting $H_0$ constraint. Thus, we present the covariance matrix of $m'_b$ as follows:

\begin{equation}
    \textbf{Cov} (\bm m'_b)=\textbf{Cov}(\bm m_b)+(5\sigma_{a{_B}})^2\bm I,
\end{equation}
where {$\bf I$} is the $n$-order square matrix where all components are all equal to $1$, and $n=1701$. Note that all quantities in bold represent vectors or matrices.

The covariance of the luminosity distances is calculated using the matrix transformation relation:

\begin{equation}\label{error_unanchored_luminosity}
\textbf{Cov}(\bm{ \Theta }^{SNe}) =
\left(\frac{\partial  \bm{  \Theta }^{SNe}} {\partial \bm{m'_b}}\right)\textbf{Cov}(\bm{m'_b}) \left(\frac{\partial  \bm{  \Theta   }^{SNe} }{\partial \bm{m'_b}}\right)^T,
\end{equation}
where the expression $\frac{\partial  {\bm \Theta} ^{\rm SNe} }{\partial {\bm m'_b}}$ represents the partial derivative matrix of the unanchored luminosity distance vector ${\bm \Theta}^{\rm SNe}$ with respect to the vector ${\bm m'_b}$. It is well established that errors in redshift measurements for SNe Ia are negligible. Therefore, no error bars are assigned to the variable $z$ in this paper, allowing it to be continuously varied in all Gaussian Processes conducted across the entire sample data \cite{Colaco2023gzy,Colaco2025aqp}.

\subsection{Gaussian Process Reconstruction}

We apply the GP regression method \cite{RasmussenW06} to the SNe Ia apparent magnitude sample and to the Einstein radius of the selected SGL sample, allowing us to reconstruct the functions $\bm{m'_{b}}(\bm{z})$ and $\bm{D}^{\rm{Obs}}$, respectively. The former observable is converted into the $\bm{\Theta}^{\rm{Obs}}(z)$ through Monte-Carlo sampling, taking into account the relation \ref{H0DL}. As mentioned earlier, we reconstruct these two observables independently at the same lens redshifts as the $D_{A,\Delta t}^{\rm{Obs}}(z_l,z_s) $ data from SGL. 

\begin{figure*}[htbp]
\includegraphics[width=0.497\textwidth]{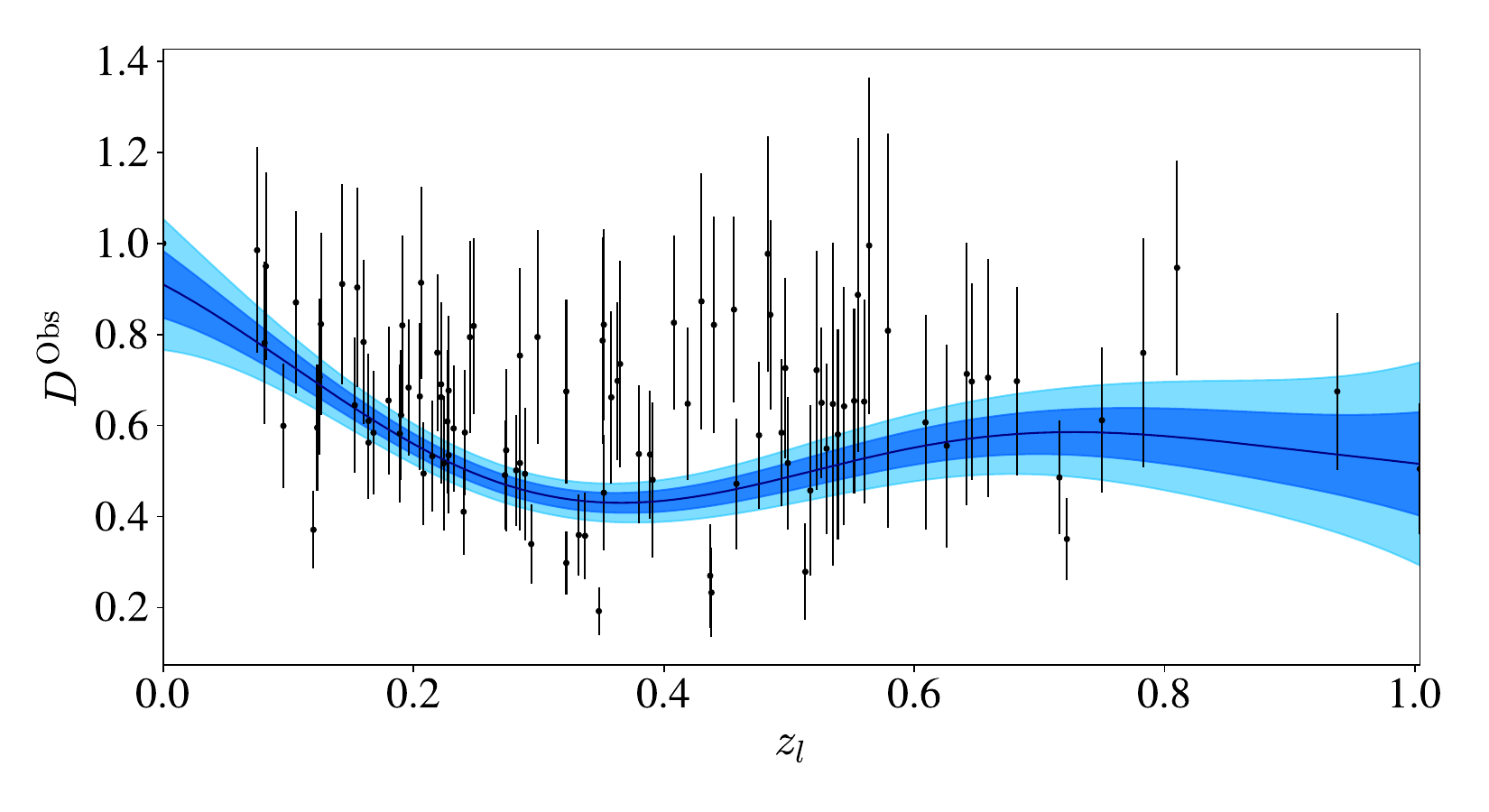}
\includegraphics[width=0.497\textwidth]{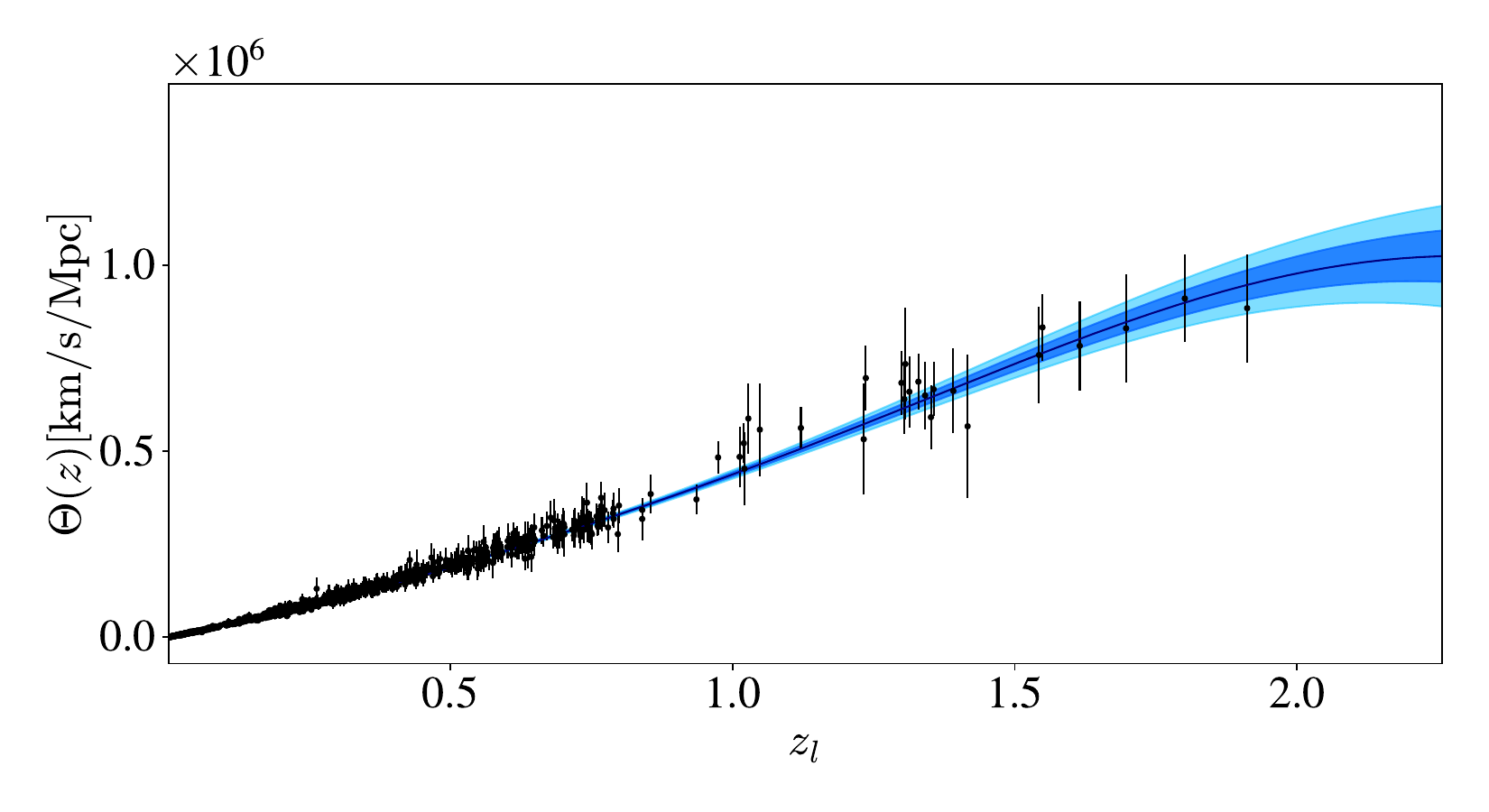}
\caption{Left Panel: The GP reconstruction of $D^{\rm{Obs}}$ regarding $z_l$ for the selected SGL systems from \cite{Cao2015qja}. Right Panel: The GP reconstruction of $\Theta^{\rm{SNe}}(z) \equiv [H_0D_L]^{\rm{SNe}}(z)$ using the SNe Ia Pantheon$+$ data compilation \cite{Scolnic2021amr}.}
\label{GPs}
\end{figure*}

This paper employs the 2.7 Python Machine Learning GaPP\footnote{Available online \url{https://github.com/carlosandrepaes/GaPP}, accessed on December 2024.} code to conduct the Gaussian Process (GP) regression \cite{2012JCAP0036S} on Type Ia supernovae and on SGL systems. The GaPP code is well-regarded for its effectiveness in machine learning tasks, as well as its user-friendliness and power. The trained network provided by the GaPP code aims to predict the unanchored luminosity distance and Einstein radius at various $z$. To achieve this, a prior mean function and a covariance function are selected. These functions quantify the correlation between the dependent variable values of the reconstruction and are characterized by a set of hyperparameters \cite{Seikel2013fda}. Typically, zero is chosen as the prior mean function to prevent biased results, and a Gaussian kernel is employed as the covariance between two data points that are separated by a redshift distance of $z - z'$, which is given by:

\begin{equation}
\label{gaussian_kernel}
k(z,z')=\sigma_f^2 \exp\left(-\frac{(z-z')^2}{2l^2}\right),
\end{equation}
where $\sigma_f$ and $l$ represent the hyperparameters that are related to the variation of the estimated function and its smoothing scale, respectively. From a Bayesian perspective, optimizing these hyperparameters typically yields a good approximation and can be computed much faster than other methods. Thus, for each observable $\mathcal{\bm{O}}_i(\bm{z})=(\bm{\Theta}^{\rm{SNe}}(\bm{z}),\bm{D}^{\rm{Obs}}(\bm{z_l}))$, it seeks to maximize the following logarithm of the marginal likelihood:

\begin{eqnarray}\label{log_likelihood}
    \ln {\mathcal{L}_{\mathcal{\bm O}}}&=&-\frac{1}{2}\mathcal{\bm O} [\bm{ K}(\bm{ z},\bm{ z})+\bm{ C}_{\mathcal{\bm{ O}}}]^{-1}( \mathcal{\bm{O}})^T \nonumber \\
&&   -\frac{1}{2}\ln{|\bm{ K}(\bm{z},\bm{ z})+ \bm{C}_{\mathcal{\bm{O}}}|}-\frac{n_d}{2}\ln{2\pi},
\end{eqnarray}
where $\bm{z}$ represents the vector of redshift measurements of each dataset. The covariance matrix, denoted as ${\bm K}({\bm z},{\bm z})$, describes the data as a GP, with its elements calculated with Eq. \ref{gaussian_kernel}, ${\bm C_{\mathcal{\bm O}}}$ is the covariance matrix of each dataset, and $n_d$ is the number of respective data points. For Pantheon$+$ dataset, $n_d = 1701$ and ${\bm C_{\mathcal{\bm O}}}$ is computed according to Eq. \ref{error_unanchored_luminosity}. For the Einstien Radius dataset, $n_d = 99$ and ${\bm{C}_{\mathcal{\bm{O}}}} = \text{diag}(\bm \sigma_{\bm{D}^{\rm{Obs}}}^{2})$ (see Figure \ref{GPs}).

As shown on the left panel of Fig. \ref{GPs}, the reconstruction of $D^{\rm{Obs}}(z_l)$ shows significant uncertainties due to the poor quality of the dataset. However, the mean value of $D^{\rm{Obs}}$ in each $z_l$ is a good approximation. By performing the GP for the current SGL dataset from \cite{Chen2018jcf}, we noted the reconstruction of $D^{\rm{Obs}}(z_l)$ showed bigger uncertainties compared to \cite{Cao2015qja} due to an unknown trend with redshift, especially at low redshifts. Although \cite{Chen2018jcf} has all systems from \cite{Cao2015qja}, we do not consider it in our analysis in order to avoid biased results. On the other hand, on the right panel of Fig. \ref{GPs}, the reconstruction of SNe Ia shows strong results at low redshifts. However, as the redshift increases, the uncertainties rise significantly due to the poor quality of data in that region. It is important to note that most kernels discussed in the literature tend to agree within the uncertainties of their predicted mean values \cite{2013MNRAS4311528B,Jesus2019nnk,Yang2015tzc,Li2021onq,Jesus2021bxq}. Therefore, when more and better data are available, it could improve future analyses, either SGL systems in low redshifts or SNe Ia observations at high redshifts, where their reconstructions currently exhibit much more significant errors.

\section{Main Results}\label{rasults}

We use Markov Chain Monte Carlo (MCMC) methods to estimate the posterior probability distribution function (pdf) of the free parameter supported by the emcee MCMC sampler \cite{2013PASP125306F}, and the GetDist Python package \cite{Lewis:2019xzd} to perform the plots. The $\chi^2$ function can be written as

\begin{equation}
    \chi^2 = (\bm{H}_0 - \bm{H}_{0,i}^{\mathrm{Est}}) \bm{C}_{\bm{H}_{0}^{\mathrm{Est}}}^{-1} (\bm{H}_0 - \bm{H}_{0,i}^{\mathrm{Est}})^{T},
\end{equation}
where $ \bm{H}_{0,i}^{\mathrm{Est}}$ are the estimates of the Hubble rate supported by Eq. \ref{H0Obs}, and $\bm{H}_0$ is a free parameter. The quantity $\bm{C}_{\bm{H}_{0}^{\mathrm{Est}}}^{-1}$ is is the inverse of the covariance matrix and it is given by $\bm{C_{H_0}} = \bm{C}_{\bm{\Theta}^{\rm{SNe}}} + \bm{C}_{\bm{D}^{\rm{Obs}}} + \bm{C}_{\bm{D_{A,\Delta t}}^{\rm Obs}}$, where $\bm{C}_{\bm{D_{A,\Delta t}}^{\rm Obs}} = \text{diag}(\bm{\sigma_{\rm{D_{A,\Delta t}^{\rm{Obs}}}}^2})$ is the diagonal matrix\footnote{Their uncertainties are not correlated.} of $D_{A,\Delta t}^{\textrm{Obs}} (z_l, z_s)$, and $\bm{C}_{\bm{\Theta}^{\rm{SNe}}}$ and $\bm{C}_{\bm{D}^{\rm{Obs}}}$ are the covariance matrixes for $\Theta^{\rm{SNe}}(z_l)$ and $D^{\rm{Obs}}(z_l)$, respectively.

The pdf is proportional to the product between the likelihood and the prior ($P(\bm{H}_0)$), that is, $P(\bm{H}_0|\bm{H}_{0}^{\mathrm{Est}})  \propto  \mathcal{L} (\bm{H}_{0}^{\mathrm{Est}}|\bm{H}_0) \cdot P(\bm{H}_0)$.

\begin{table}
\label{stats1}
    \renewcommand{\arraystretch}{1.2}
    \setlength\tabcolsep{0.1cm}%
	\begin{tabular}{|c|l|l|c|} \hline
{\bf Observational data}& {\bf $H_0$} [km/s/Mpc] & {\bf Ref.}\\  \hline 
SNe Ia +SGL $(D_{A,\Delta t}$+$\theta_E)$ & $70.55 \pm 7.435$ & \bf{This Paper} \\
SGL+GRBs& $72.90 \pm 3.70$& \cite{Du2023zsz} \\ 
BAO+CC+SNe Ia& $69.50 \pm 1.70$& \cite{Renzi2020fnx} \\
QSO+SGL+SNe Ia& $70.80 \pm 1.50$& \cite{Li2023gpp} \\
ESZ+S$_X$&$67.22 \pm 6.07$&\cite{Colaco2023gzy} \\
$f_{\textrm{gas}}$+SNe Ia& $73.40 \pm 5.95$& \cite{Gonzalez2024qjs} \\
SNe + $D_{A,\Delta t}(z_l,z_s)$ & $75.57 \pm 4.415$ & \cite{Colaco2025aqp} \\ 
SNe Ia+QSO&$73.50 \pm 0.67$& \cite{Liu2023ifz} \\
BAO+SNe+$H(z)$&$68.60 \pm 2.50$& \cite{Renzi2021xii} \\ \hline
	\end{tabular}
	\caption{Recent $H_0$ estimates by using Eq. \ref{H0NA} with distinct observational data.}
\end{table}

\begin{figure}[htbp]
\includegraphics[width=0.5\textwidth]{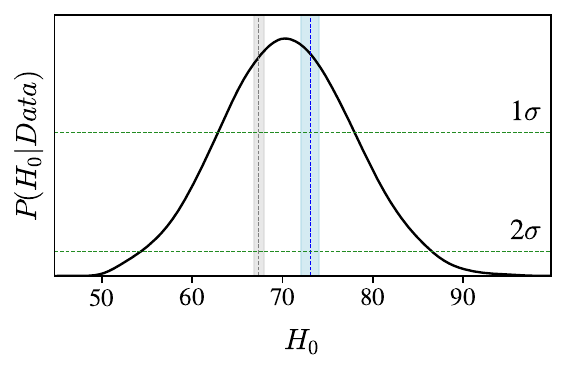}
\caption{The pdf of $H_0$ with additional intrinsic error. The light green horizontal dashed lines correspond to $1\sigma$ and $2\sigma$ confidence levels, and the blue and grey vertical dashed lines correspond to the Planck collaboration (2018) \cite{Planck2018vyg} and Riess \textit{et al}. (2020) \cite{Riess2021jrx} estimates with corresponding 1$\sigma$ confidence region, respectively.}
\label{Statistics}
\end{figure}

We obtain at $68\%$ c.l. the following result (see Fig. \ref{Statistics}): $H_0 = 70.55 \pm 7.435$ km/s/Mpc with $\sigma_{\textrm{int}} \approx 24 \%$. This intrinsic uncertainty of approximately $\sigma_{\textrm{int}} \approx 24 \%$ is included in $C_{\bm{H_{0}^{\rm{Est}}}}^{-1}$ to take into account for possible random deviations from the PLAW model and is necessary to obtain a $\chi_{red}\approx 1$. For comparison, we also show the $68\%$ regions of the estimates from the Planck Collaboration (2018) \cite{Planck2018vyg}, represented by the grey dashed vertical line, and from Riess \textit{et al}. (2020) \cite{Riess2021jrx}, represented by the blue dashed vertical line. As illustrated in Fig. \ref{Statistics}, our result falls between the Planck and SH0ES values at the $1\sigma$ region. This apparent consistency with both values is meaningful given the large uncertainties of our estimate. Although our result does not provide a new resolution to the current Hubble tension, it demonstrates the robustness and potential of combining strong lensing and supernova data through the CDDR to constrain $H_0$ in a model-independent way. This methodology establishes a solid foundation for future efforts at achieving more precise and reliable determinations of $H_0$, especially as larger and higher-quality samples of lensing systems become available. 

Table I presents our main result alongside other results that used Eq. \ref{H0NA} to infer $H_0$. As one may see, the value obtained here by combining strong lensing and SNe Ia data is competitive with other recent results that employed the same methodology. However, the uncertainties in our estimate are relatively large, as they are based on only seven Gravitational lens time-delay systems, which limits the statistical power and increases the sensitivity to systematic effects.

It is important to note that the determination of $H_0$ has exhibited variability across different analyses of gravitational lensing systems. The H0LiCOW Collaboration inferred $H_0 = 73.3^{+1.7}_{-1.8}~\mathrm{km\,s^{-1}\,Mpc^{-1}}$ using strong gravitational lenses, based on the assumption of a flat $\Lambda$CDM model. Subsequent analyses by the TDCOSMO Collaboration yielded a range of values for $H_0$, which includes $H_0 = 74.2^{+1.6}_{-1.6}~\mathrm{km\,s^{-1}\,Mpc^{-1}}$ based on a power-law mass profile, and $H_0 = 67.4^{+4.3}_{-4.7}~\mathrm{km\,s^{-1}\,Mpc^{-1}}$ when incorporating data from the SLACS survey (for details, see Ref. \cite{Birrer2020tax}). The observed discrepancies between these values reflect the complexity of cosmological measurements and the varying assumptions made in modeling the mass profiles of lensing galaxies. While the value obtained using the power-law profile provides a higher and more restrictive estimate, the value derived from SLACS observations, which includes kinematic data and more stringent considerations of mass distribution and anisotropy, yields a significantly lower and less restrictive estimate. This difference underscores the importance of accounting for dynamical variables and the internal structure of galaxies when estimating $H_0$, emphasizing the impact of different methodologies and assumptions on cosmological conclusions (see also \cite{2006glsw.conf.....M,2024MNRAS.533..795K,2016JCAP...08..020B,Birrer2020tax}). In light of this,  we clarify that our analysis relies on stellar velocity dispersion measurements from the H0LiCOW Collaboration to constrain the mass models of the lens systems. These measurements are interpreted using a simplified dynamical model, which generally assumes isotropic or mildly anisotropic stellar orbits. Since our analysis does not explicitly marginalize over anisotropy models, the quoted $H_0$ should be interpreted within the context of this limitation. Nonetheless, the technique presented here remains robust due to its independence from any specific cosmological model, and it holds significant potential for yielding more accurate estimates in future applications that incorporate more realistic and detailed mass profiles.

\section{Final Remarks}
\label{final}

In this work, we have presented a cosmology model-independent determination of the Hubble constant $H_0$ by combining observations of strong gravitational lensing systems with Type Ia Supernovae data. Using seven lens systems from the TDCOSMO Collaboration, we calculated the angular diameter distance to each lens $D_{A_l}$ using the product $D^{\textrm{Obs}}(z_l) \cdot D_{A,\Delta t}^{\textrm{Obs}}(z_l, z_s)$. The quantity $D^{\textrm{Obs}}(z_l)$ was reconstructed at each lens redshift $z_l$ using Gaussian Process regression techniques, applied to a dataset of Einstein radius measurements. Meanwhile, $D_{A,\Delta t}^{\textrm{Obs}}(z_l, z_s)$ represents the time-delay angular distance for each system.

We also reconstructed the unanchored luminosity distance ($\Theta^{\textrm{SNe}}(z) \equiv [H_0 D_L]^{\textrm{SNe}} (z)$) at the same redshifts using Gaussian Process techniques applied to the Pantheon+ Type Ia Supernovae dataset, which includes the full apparent magnitude covariance matrix. Connecting these measurements through the validity of the distance duality relation (see Eq. \ref{H0Obs}), we found a value of $H_0 = 70.55 \pm 7.435$ km/s/Mpc at $68\%$ c.l..  Since our analysis does not explicitly marginalize over anisotropy models, our quoted $H_0$ should be interpreted within the context of this limitation.  From Fig. \ref{Statistics}, it is possible to verify that the $H_0$ estimate from the Planck Collaboration (2018) and from Riess \textit{et al}. (2020) fall within our analysis's $68\%$ confidence region. Such an apparent consistency with both values is meaningful given the large uncertainties of our estimate. While our result does not yet shed new light on the current Hubble tension, it represents an important, fully model-independent consistency test that agrees with both local and early-Universe determinations within the uncertainties. More importantly, it illustrates the viability of combining strong lensing and supernova data via the distance duality relation to constrain $H_0$ without relying on a cosmological model.

As is well known, the measurement of $H_0$ using gravitational lensing observations is deeply limited by the present mass-sheet degeneracy (MSD) in lens mass modeling. This significant issue (see \cite{2006glsw.conf.....M,2024MNRAS.533..795K,2016JCAP...08..020B}) introduces an inherent uncertainty in determining the lens potential, leading to biased estimates of $H_0$. To overcome this challenge, future analyses must mitigate the impact of this degeneracy by integrating additional observational constraints, such as high-resolution imaging, stellar kinematics, and refined mass profile models. The forthcoming generation of Telescopes \textemdash including the Euclid mission, the Nancy Grace Roman Space Telescope, and the Vera C. Rubin Observatory \textemdash will significantly increase the sample sizes, enhancing statistical power in this area. Moreover, the James Webb Space Telescope will play a crucial role in advancing this effort (see \cite{Treu2022aqp} for more information). Therefore, the approach presented here paves the way for more precise and robust constraints on the Hubble constant with upcoming larger samples of lensing systems.

\vspace{1cm}

\begin{acknowledgments}

\noindent The authors thank Javier E. Gonzalez for his valuable contribution to this manuscript regarding discussions over Gaussian Process. LRC thanks the financial support from the Conselho Nacional de Desenvolvimento CientÍfico e Tecnológico (CNPq, National Council for Scientific and Technological Development) under the project No. 169625/2023-0. RFLH also thanks CNPq for support under project No. 309132/2020-7.

\end{acknowledgments}

\bibliography{PRD}

\end{document}